\documentclass[twocolumn,preprintnumbers,nofootinbib]{revtex4}
\usepackage{amsmath} \usepackage{graphicx} \usepackage{amsfonts}
\usepackage{array} \usepackage{amsthm} \usepackage{bm} \usepackage{mathptmx}

\usepackage{latexsym}
\usepackage{hyperref}
\hypersetup{colorlinks=false}

\newcommand{\nn}{\nonumber}
\newcommand{\be}{\begin{equation}}
\newcommand{\ee}{\end{equation}}
\newcommand{\ba}{\begin{eqnarray}}
\newcommand{\ea}{\end{eqnarray}}
\newcommand{\bal}{\begin{align}}
\newcommand{\eal}{\end{align}}

\newcommand{\e}{{\rm e}}
\newcommand{\dd}{{\rm d}}
\newcommand{\ii}{{\rm i}}
\newcommand{\bb}{\bibitem}

\newcommand{\om}{\omega}

\newcommand{\ro}{\rho}
\newcommand{\ep}{\epsilon}

\newcommand{\ta}{\theta}

\newcommand{\Si}{\Sigma}
\newcommand{\De}{\Delta}

\newcommand{\Om}{\Omega}

\newcommand{\bw}{\begin{widetext}}
\newcommand{\ew}{\end{widetext}}

\def\abh{black hole }
\def\bh{black holes }
\def\aBH{black hole}
\def\BH{black holes}

\def\Sd{Schwarzschild }

\begin{document}
\title{From static to rotating to conformal static solutions: Rotating imperfect fluid wormholes \\with(out) electric or magnetic field}
\author{Mustapha Azreg-A\"{\i}nou}
\affiliation {Ba\c{s}kent University, Department of Mathematics,
Ba\u{g}l\i ca Campus, Ankara, Turkey}


\begin{abstract}
We derive a shortcut stationary metric formula for generating imperfect fluid rotating solutions, in Boyer-Lindquist coordinates, from spherically symmetric static ones. We explore the properties of the curvature scalar and stress-energy tensor for all types of rotating regular solutions we can generate without restricting ourselves to specific examples of regular solutions (regular black holes or wormholes). We show through examples how it is generally possible to generate an imperfect fluid regular rotating solution via radial coordinate transformations. We derive rotating wormholes that are modeled as imperfect fluids and discuss their physical properties that are independent on the way the stress-energy tensor is interpreted. A solution modeling an imperfect fluid rotating loop black hole is briefly discussed. We then specialize to the recently discussed stable exotic dust Ellis wormhole emerged in a source-free radial electric or magnetic field, generate its, conjecturally stable, rotating counterpart which turns out to be an exotic imperfect fluid wormhole and determine the stress-energy tensor of both the imperfect fluid and the electric or magnetic field.

\end{abstract}


\maketitle

\section{Introduction\label{sec0}}

Applications of rotating solutions to astrophysics and theories of gravity are of great importance. Many of the solutions derived in this context are linear approximations with respect to the rotating parameter $a$ or angular momentum $J$~\cite{as1}-~\cite{gh}. When the linear approximation is no longer valid, as is the case with fast rotating objects in the cosmos, only the well known set of exact solutions~\cite{Van}-~\cite{Kramer} (and references therein) has been, or may be, used for matching exterior vacuum configurations to interior fluid cores~\cite{Herrera}-~\cite{Lemos}, and references therein.

Generating rotating solutions by linearization does not generally demand a special approach but appeals to symmetry properties~\cite{as1}-~\cite{gh}. In contrast, most of the approaches used to derive exact rotating solutions, besides relying on symmetry properties,  were methodic~\cite{Van}-~\cite{Kramer} and~\cite{Carter}-~\cite{phantom1} or partly methodic relying on some ad hoc hypotheses~\cite{Janis1}-~\cite{Janis3}.

The Newman-Janis algorithm (NJA)~\cite{Janis1} was first devised to generate exterior rotating solutions but later was applied to generate rotating interior metrics which were matched to the exterior Kerr one~\cite{Herrera,Via}. The metric we intend to derive has the property to generate both interior and exterior rotating solutions in Boyer-Lindquist coordinates (BLC's) and it avoids the ambiguous complexification procedure. Since the NJA is well known, we will skip details about its application (see, for instance,~\cite{Janis2,Janis4}).

In Sect.~\ref{sec1} we derive the stationary metric for generating rotating solutions. Sect.~\ref{sec2} is devoted to a general discussion of the properties of the curvature scalar and stress-energy tensor for all types of regular rotating solutions we can generate without restricting ourselves to specific examples of regular static solutions (regular black holes or wormholes). Explicit examples of rotating imperfect fluid wormholes and loop \bh are provided and briefly discussed. In Sect.~\ref{sec3}, we apply the rules and derive rotating wormholes that are modeled as imperfect fluids and discuss their physical properties that are independent on the way the stress-energy tensor is interpreted. We then specialize to the recently discussed stable exotic dust Ellis wormhole emerged in a source-free radial electric or magnetic field, generate its, conjecturally stable, rotating counterpart which turns out to be an exotic imperfect fluid wormhole and determine the stress-energy tensor of both the imperfect fluid and the electric or magnetic field. Our concluding remarks are made in Sect.~\ref{sec4}. An appendix has been added to prove uniqueness of some solutions and to ease the discussion made in Sect.~\ref{sec2}.

\section{The rotating metric\label{sec1}}

Consider the static metric\footnote{It is always possible, by a coordinate transformation $r\to R(r)$, to bring~\eqref{1} to the form where the transformed function $F=G$ but it may not be possible to express $H$ in terms of $R$ as is the case with some wormhole solutions and regular \BH.}
\begin{equation}\label{1}
\dd s_{\text{stat}}^2 = G(r)\dd t^2 - \frac{\dd r^2}{F(r)} - H(r)(\dd \ta^2+\sin^2\ta\dd \varphi^2)
\end{equation}
to which we apply, in a first step, the NJA. For the sake of subsequent applications (to regular \bh and wormholes), we will not assume $H= r^2$ nor will we assume $G= F$. After introducing the advanced null coordinates ($u,r,\ta,\varphi$) defined by $\dd u=\dd t-\dd r/\sqrt{FG}$, the nonzero components of the resulting inverse metric are of the form $g^{\mu \nu }=l^{\mu} n^{\nu} +  l^{\nu} n^{\mu} -  m^{\mu} {\bar m}^{\nu} -  m^{\nu} {\bar m}^{\mu}$ with
\begin{align}
& l^{\mu } = \delta_r^{\mu} \nonumber \\
\label{2}& n^{\mu} = \sqrt{F/G} \,\delta^{\mu}_u - (F/2) \delta^{\mu}_r  \\
& m^{\mu} = \big( \delta_{\theta}^{\mu} + \frac{ \ii}{\sin \theta}\,\delta_{\varphi}^{\mu} \big)/\sqrt{ 2 H}\nonumber
\end{align}
and $l_{\mu} l^{\mu} = m_{\mu} m^{\mu} = n_{\mu} n^{\mu} = l_{\mu} m^{\mu} =n_{\mu} m^{\mu} =0$
and $l_{\mu} n^{\mu} = - m_{\mu} {\bar m}^{\mu} =1$. Now, if we perform the complex transformation
\begin{equation}\label{3}
r \to r + \ii a \cos \theta,\;u \to  u  - \ii a \cos \theta
\end{equation}
then, $\delta_{\nu}^{\mu}$, if treated as vectors, transform as $\delta_r^{\mu}\to\delta_r^{\mu}$, $\delta_u^{\mu}\to\delta_u^{\mu}$, $\delta_{\theta}^{\mu}\to\delta_{\theta}^{\mu}+\ii a\sin\ta(\delta_u^{\mu}-\delta_r^{\mu})$ and $\delta_{\varphi}^{\mu}\to\delta_{\varphi}^{\mu}$, and we assume that \{$G(r),F(r),H(r)$\} transform to \{$A(r,\ta,a),B(r,\ta,a),$ $\Psi(r,\ta,a)$\} where \{$A,B,\Psi$\} are three-variable real functions, to be fixed later\footnote{We may subject them to the constraints
\begin{multline}\label{4}
\lim_{a\to 0}A(r,\ta,a)=G(r),\;\;\lim_{a\to 0}B(r,\ta,a)=F(r),\\
\lim_{a\to 0}\Psi(r,\ta,a)=H(r),
\end{multline}
if we want to recover~\eqref{1} in the limit $a\to 0$. However, these constraints are nonrestrictive and we may drop them as we did in~\cite{conf}. When this is the case, the limit $a\to 0$, in the rotating metric to be derived [Eq.~\eqref{10}], leads to a static metric conformal to~\eqref{1}.}.

The effect of the transformation~\eqref{3} on ($l^{\mu },n^{\mu},m^{\mu}$) is the `product' of the transformations on $\delta_{\nu}^{\mu}$ and \{$G(r),F(r),\\H(r)$\}:
\begin{align}
& l^{\mu } = \delta_r^{\mu} \nonumber \\
\label{5}& n^{\mu} = \sqrt{B/A} \,\delta^{\mu}_u - (B/2) \delta^{\mu}_r  \\
& m^{\mu} = \big[ \delta_{\theta}^{\mu}+\ii a\sin\ta(\delta_u^{\mu}-\delta_r^{\mu}) + \frac{ \ii}{\sin \theta}\,\delta_{\varphi}^{\mu} \big]/\sqrt{ 2 \Psi}.\nonumber
\end{align}

By imposing no constraints on \{$A,B,\Psi$\} --even if we impose~\eqref{4}-- we assert that our approach and the usual NJA differ starting from this step, that is, we do not assume that \{$A,B,\Psi$\} are derived from \{$G,F,H$\} by some sort(s) of complexification of $r$ which is an ambiguous procedure, not unique, and leads to nonphysical solutions~\cite{kr} that cannot be written in BLC's as shown in~\cite{az2}. Rather, we impose the requirement that the final rotating metric be written in BLC's which, as we shall see below, fixes uniquely the functions \{$A,B$\}. The determination of $\Psi$ depends on the physical problem at hands, that is, it depends on the type of rotating solution one wants to derive. $\Psi$ generally obeys some partial differential equation(s). In the case to which one is generally interested, where the source term in the field equations, $T^{\mu\nu}$, is interpreted as an imperfect fluid, these partial differential equations are given below [Eqs.~\eqref{13}, \eqref{13r}]. Thus, the essence of our procedure is to reduce the task of determining the rotating counterpart of~\eqref{1} to that of fixing $\Psi$ by solving nonlinear partial differential equations where `nonlinearity' results in different rotating solutions for a given static one. Applications are considered in Sects.~\ref{sec2} and~\ref{sec3}. Some other applications are found in~\cite{conf}.

Knowing the transformed vectors~\eqref{5}, we obtain the transformed inverse metric
\begin{align}
& g^{u u }(r,\ta) = -\frac{a^2 \sin^2\theta}{ \Psi}, \;\;
g^{u \varphi}(r,\ta) = -\frac{a}{ \Psi} , \nonumber  \\
& g^{\varphi \varphi}(r,\ta) = -\frac{1}{\Psi\sin^2 \theta}  , \;\;
g^{\theta \theta}(r,\ta) = -\frac{1}{ \Psi},  \nonumber \\
& g^{rr}(r,\ta) = -B-\frac{a^2 \sin ^2 \theta }{ \Psi} , \;\;
g^{r \varphi}(r,\ta) = \frac{a}{\Psi} ,\nonumber  \\
\label{6}& g^{u r }(r,\ta) = \sqrt{\frac{B}{A}}+\frac{a^2 \sin ^2\theta }{\Psi},
\end{align}
and then the rotating metric in Eddington-Finkelstein coordinates
\begin{multline}\label{7}
\dd s^2 = A \dd u^2+2 \frac{\sqrt{A}}{\sqrt{B}} \dd u \dd r+2 a\sin ^2\theta  \Big(\frac{\sqrt{A}}{\sqrt{B}}-A\Big) \dd u \dd \varphi \\- 2 a\sin ^2\theta  \frac{\sqrt{A}}{\sqrt{B}} \dd r \dd \varphi-\Psi  \dd \theta^2 \\- \sin ^2\theta  \Big[\Psi +a^2\sin ^2\theta  \Big(2 \frac{\sqrt{A}}{\sqrt{B}}-A\Big)\Big] \dd \varphi^2.
\end{multline}
Setting
\begin{equation}\label{K}
K(r)\equiv \sqrt{F}H/\sqrt{G},
\end{equation}
the metric~\eqref{7} is brought to BLC's on performing the coordinate transformation
\begin{equation}\label{8}
\dd u=\dd t - \frac{(K+a^2)\dd r}{FH+a^2},\,\dd\varphi=\dd\phi -\frac{a\dd r}{FH+a^2}
\end{equation}
provided we choose
\begin{equation}\label{9}
A(r,\ta)=\frac{(FH+a^2\cos^2\ta) \Psi }{(K+a^2 \cos^2\ta)^2},\, B(r,\ta)=\frac{FH+a^2\cos^2\ta}{\Psi}.
\end{equation}
Finally, the desired form of the rotating solution is
\begin{align}
&\dd s^2 = \frac{(F H+a^2 \cos ^2\theta ) \Psi \dd t^2}{(K+a^2 \cos ^2\theta )^2}
    -\frac{\Psi \dd r^2}{F H+a^2}\nn\\
&+2 a \sin ^2\theta\Big[\frac{K- FH}{(K+a^2 \cos ^2\theta )^2}\Big]\Psi\dd t\dd \phi -\Psi\dd \ta^2\nonumber\\
\label{10}&-\Psi\sin ^2\theta\Big[1+a^2\sin ^2\theta\frac{2K- FH
    +a^2\cos ^2\theta }{(K+a^2\cos ^2\theta )^2}\Big]\dd \phi^2.
\end{align}
Setting $\ro^2\equiv K+a^2 \cos^2\ta$, $2f(r)\equiv K-FH$, $\De(r)\equiv FH+a^2$ and $\Si\equiv (K+a^2)^2-a^2\De\sin^2\ta$ we bring~\eqref{10} to the following useful Kerr-like metrics
\begin{align}
&\dd s^2 =\frac{\Psi}{\ro^2}\Big[\Big(1-\frac{2f}{\ro^2}\Big)\dd t^2-\frac{\ro^2}{\De}\,\dd r^2\nn\\
\label{10a}&+\frac{4af \sin ^2\theta}{\ro^2}\,\dd t\dd \phi-\ro^2\dd \ta^2-\frac{\Si\sin ^2\theta}{\ro^2}\,\dd \phi^2\Big]
\end{align}
\begin{align}
\label{10b}&\dd s^2 =\frac{\Psi}{\ro^2}\Big[\frac{\De}{\ro^2}(\dd t-a\sin^2\ta\dd \phi)^2-\frac{\ro^2}{\De}\,\dd r^2-\ro^2\dd \ta^2\nn\\
&-\frac{\sin^2\ta}{\ro^2}\,[a\dd t-(K+a^2)\dd \phi]^2\Big].
\end{align}

A generalization of~\eqref{10} is possible on modifying the complex transformation~\eqref{3}.

For \emph{fluid} solutions that rotate about the $z$ axis, we fix $\Psi(r,\ta,a)$ upon solving the field equation $G_{r\ta}\equiv 0$. As we shall see later another constraint will be imposed on $\Psi$ to ensure consistency of the field equations, $G_{\mu\nu}=T_{\mu\nu}$, for the form of the fluid source term we will work with.

Due to its nonlinearity, $G_{r\ta}\equiv 0$ possesses different solutions~\cite{conf}. For given \{$G,F,H$\}, those solutions $\Psi_n$ which obey the extra constraints~\eqref{4} have been called normal fluids, and those $\Psi_c$ which do not obey them have been called conformal fluids~\cite{conf}, their metrics are conformally related
\begin{equation}\label{12}
    \dd s_c^2=(\Psi_c/\Psi_n)\dd s_n^2.
\end{equation}
As discussed in~\cite{conf}, conformal fluids have more interesting properties than normal ones and may also be used as interior regular cores. Now, since $\lim_{a\to 0}\Psi_c\neq H$ (by definition) and $\lim_{a\to 0}\dd s_n^2= \dd s_{\text{stat}}^2$ [Eq.~\eqref{1}], this implies that $\lim_{a\to 0}\dd s_c^2\neq \dd s_{\text{stat}}^2$. Hence, $\lim_{a\to 0}\dd s_c^2$ is a new static metric conformal to $\dd s_{\text{stat}}^2$. Conversely, had we started from the static solution $\lim_{a\to 0}\dd s_c^2$ we would have recovered $\dd s_{\text{stat}}^2$ from the limit $a\to 0$ of Eq.~\eqref{10}, taking $\Psi=\Psi_n$ in this latter equation. This is obvious because, setting $\lim_{a\to 0}(\Psi_c/\Psi_n)=C(r)$, the transformation \{$G,F,H$\} $\leftrightarrow$ \{$CG,F/C,CH$\} keeps invariant $\dd s^2/\Psi$ in~\eqref{10}. This in return implies that the two fluids are dual to each other.

\section{The curvature scalar and stress-energy tensor\label{sec2}}

Due to symmetry properties, each metric component in~\eqref{10} must be an even function of $a$, except the mixed term which must be odd, this implies that $\Psi$ is an even function of $a$. It is then more convenient to look for solutions of the form $\Psi\equiv \Psi(r,y^2,a^2)$ where $y\equiv \cos\ta$. Introducing an indexical notation for derivatives: $\Psi_{,ry^2}\equiv \partial^2\Psi/\partial r\partial y^2$, $K_{,r}\equiv \partial K/\partial r$, etc, the equation $G_{r\ta}\equiv 0$ yields
\begin{equation}\label{13}
(K+a^2 y^2)^2 (3\Psi_{,r}\Psi_{,y^2} -2\Psi \Psi_{,ry^2}) =3a^2K_{,r}\,\Psi^2.
\end{equation}

We work with an orthonormal basis $(e_t,\,e_r,\,e_{\ta},\,e_{\phi})$ which is dual to the 1-forms defined in~\eqref{10b}: $\om^t\equiv \sqrt{\Psi\De}(\dd t-a\sin^2\ta\dd \phi)/\ro^2$, $\om^r\equiv -\sqrt{\Psi}\dd r/\sqrt{\De}$, $\om^{\ta}\equiv -\sqrt{\Psi}\dd \ta$, $\om^{\phi}\equiv -\sqrt{\Psi}\sin\ta [a\dd t-(K+a^2)\dd \phi]/\ro^2$:
\begin{align}
&e^{\mu}_t=\frac{(K+a^2,0,0,a)}{\sqrt{\Psi \De}}\,,\;e^{\mu}_r=\frac{\sqrt{\De}(0,1,0,0)}{\sqrt{\Psi}}\nn\\
\label{16}&e^{\mu}_{\ta}=\frac{(0,0,1,0)}{\sqrt{\Psi}}\,,\;e^{\mu}_{\phi}=-\frac{(a\sin^2\ta,0,0,1)}{\sqrt{\Psi}\sin\ta}.
\end{align}
where $e^{\mu}_t$ is the 4-velocity vector of the fluid. With $G_{r\ta}\equiv 0$, the source term may be represented as an imperfect fluid whose SET is of the form
\begin{equation}\label{17}
    T^{\mu\nu}=\ep e^{\mu}_te^{\nu}_t+p_re^{\mu}_re^{\nu}_r+p_{\ta}e^{\mu}_{\ta}e^{\nu}_{\ta}+p_{\phi}e^{\mu}_{\phi}e^{\nu}_{\phi}
\end{equation}
where $\ep$ is the density and ($p_r,\,p_{\ta},\,p_{\phi}$) are the components of the pressure. As we shall see in Sect.~\ref{sec3b}, other representations are possible. A consistency check of the field equations $G_{\mu\nu}=T_{\mu\nu}$ and the form of $T_{\mu\nu}$, Eq.~\eqref{17}, yields the linear partial differential equation
\begin{multline}\label{13r}
\Psi[{K_{,r}}^2+K(2-K_{,rr})-a^2y^2(2+K_{,rr})]\\+(K+a^2y^2)(4y^2\Psi_{,y^2}-K_{,r}\Psi_{,r})=0.
\end{multline}

Among solutions to the system~\eqref{13} and~\eqref{13r} of the form $\Psi\equiv g(\ro^2)$, we have shown that the special solution~\cite{conf}
\begin{equation}\label{14}
    \Psi_s=r^2+q^2+a^2y^2,\quad (K=r^2+q^2)
\end{equation}
is unique up to a multiplicative constant (which is conformal if $G\neq F$, that is if $H\neq K=r^2+q^2$; otherwise it is normal). Here $q^2$ is a real constant. Moreover, it is also possible to show that~\eqref{14} is the unique power-law solution of the form $[l(r)+a^2y^2 k(r)]^m$. Hence, the hope to find a simple solution obeying~\eqref{4}, that is where $l(r)=H(r)$, vanishes. Other solutions than $\Psi_s$ that may obey~\eqref{4} have thus more complicated structures which we write as:
\begin{equation}\label{15}
    \Psi_n=H\exp{[a^2\psi(r,y^2,a^2)]},\;(\lim_{a\to 0}a^2\psi=0).
\end{equation}
Solutions of the form~\eqref{15} have Taylor expansions in powers of $a^2$ of the form $\Psi_n=H+\sum_{i=1}a^{2i}X_{2i}(r,y^2)$ where the first term, or independent term, of the series is $H$. It is shown in the appendix that if $G=F$ ($K=H$), $\Psi_s$, with $H=K=r^2+q^2$, is the unique solution of this type~\eqref{15}. If $G\neq F$ ($K\neq H$), $\Psi_s$ is no longer of the form~\eqref{15} (see next paragraph); however, other solutions of the form~\eqref{15} exist in this case too.

Note that any general solution $\Psi_g$ to~\eqref{13}, \eqref{13r} may be brought to the form~\eqref{15} but without the extra condition $\lim_{a\to 0}a^2\psi=0$:
\begin{equation}\label{15rr}
    \Psi_g=H\exp{(a^2\psi)}.
\end{equation}
For instance one can write $\Psi$ of the form: $\Psi_s=H\exp{(a^2\psi)}$ with $a^2\psi=\ln[(K+a^2y^2)/H]$ and $\lim_{a\to 0}a^2\psi =\ln(K/H)$. One sees that $\Psi_s$ is normal, of the form~\eqref{15}, only in the case $H=K$.

In the case $G=F$, the Kerr and the rotating de Sitter solution were derived in~\cite{conf} and examples of normal and conformal regular rotating cores were given too. It is straightforward to use~\eqref{10a} to derive regular rotating \bh from each known regular static one~\cite{regular}: All one needs is to insert the metric \{$G,F,H$\} of the static regular hole in~\eqref{10a} along with $\Psi=\Psi_s$. To our knowledge, existing static regular \bh have \{$G,F,H$\}=\{$F,F,r^2$\}, yielding $K=H=r^2$ and $q^2=0$ [Eq.~\eqref{14}], so that $\Psi_s=r^2+a^2y^2$ and all derived regular rotating \bh will be normal. However, we won't do that here since, after constructing~\eqref{10}, our second purpose is to extend the analysis to rotating fluid wormholes and we will include a discussion on rotating fluid loop \BH. A part of the application of~\eqref{10a} is given in this section and the other part is postponed to Sect.~\ref{sec3}.

In the remaining part of this section, we will investigate the properties of the curvature scalar $R$ and stress-energy tensor (SET) $T^{\mu\nu}$ for all types of regular rotating solutions we can derive using~\eqref{10} or~\eqref{10a}, taking $\Psi=\Psi_s$ or $\Psi=\Psi_g$ as defined in~\eqref{14} and~\eqref{15r} without restricting ourselves to specific examples of regular static solutions and, unless otherwise specified, we assume $G\neq F$. We will at the same time provide explicit examples of rotating imperfect fluid wormholes and loop \bh and give instances of the possibility to generate simple imperfect (conformal or normal) fluid rotating solution to any given static one via a radial coordinate transformation $r\to R(r)$. Other examples were given in~\cite{conf}.

Using thus $\Psi_g$ as a general form of any solution to~\eqref{13}, \eqref{13r} we derive the components of the SET from the field equations $G_{\mu\nu}=T_{\mu\nu}$ by
\begin{align}
&\epsilon =\frac{1}{\Psi_g }-\frac{a^2 [20 y^2 (K+a^2)+24 y^2 f+(1-y^2) K_{,r}^{\ \; 2}] }{4 \Psi_g  \rho ^4}\nn\\
&+\frac{3 \Delta  (H_{,r}+a^2 H \psi _{,r})^2-4 a^4 y^2 (1-y^2)
H^2 \psi _{,y^2}^{\ \ 2}}{4 H^2 \Psi_g }\nn\\
&+\frac{2 a^2}{\Psi_g\rho ^2}+\frac{2 a^2 [a^2 y^2 (1+y^2)-(1-3 y^2) K] \psi _{,y^2}}{\Psi_g  \rho ^2}\nn\\
&-\frac{1}{2 H \Psi_g } \{8 a^2 y^2 (1-y^2) H \psi _{,y^2y^2} +\Delta _{,r} (H_{,r}+a^2H \psi _{,r})\nn\\
\label{18a}&+2 \Delta  [H_{,rr}+a^2 [2 H_{,r} \psi _{,r}+H (a^2 \psi _{,r}^{\ \; 2}+\psi _{,rr} )]]\}
\end{align}
\begin{align}
&p_r=-\ep +\frac{2 a^2y^2\Delta}{\Psi_g  \rho ^4}-\frac{\Delta  (H_{,r} K_{,r}+a^2 H K_{,r} \psi _{,r})}{H \Psi_g  \rho ^2}\nn\\
&+\frac{\Delta}{2 H^2 \Psi_g}[3 H_{,r}^{\ \; 2}-2 H H_{,rr}+2 a^2 H H_{,r}
\psi _{,r}\nn\\
\label{18b}&+a^4 H^2 \psi _{,r}^{\ \; 2}-2 a^2 H^2 \psi _{,rr}]
\end{align}
\begin{align}
&p_{\ta}-p_{\phi}=\frac{a^2 (1-y^2) K_{,r}^{\ 2}}{2 \Psi_g  \rho ^4}-\frac{4 a^4 (1-y^2) y^2 \psi _{,y^2}}{\Psi_g  \rho ^2}\nn\\
\label{18c}&+\frac{2 a^2 (1-y^2) (a^2 y^2 \psi _{,y^2}^{\ \ 2}-2 y^2 \psi_{,y^2y^2}-\psi _{,y^2})}{\Psi_g }.
\end{align}
If $\Psi=\Psi_s$, these expressions reduce to Eqs.~(13), (14) of~\cite{conf} in case $G=F$ or to Eqs.~(18), (19) of~\cite{conf} in case $G\neq F$.

The general expression of the curvature scalar $R=N/D$, where $D\equiv 2\ro^4\Psi_g^3$ and $N$ is a polynomial in ($\ro^2,y^2$) and in ($K,F,H,\Psi_g$) and their first and second order derivatives, may be simplified further if $N$ has common factors with $D$. From now on, $N$ and $D$ denote the simplified numerator and denominator of $R$. Depending on the nature of the static solution~\eqref{1} (regular \abh or wormhole), the ring singularity $\ro^2=0$, if any, may occur at $r=0$ [$K(0)=0$, $F(0)\neq 0$ and $H(0)=0$] if the solution is a \abh or at the throat $r=r_{\text{th}}>0$, which is defined by $H(r_{\text{th}})=r_0^2 > 0$ [$K(r_{\text{th}})=0$ and $F(r_{\text{th}})=0$], if the solution is a wormhole. Here $r_0^2$ is the minimum value of $H(r)$.\\

\paragraph*{\textbf{Case (1):} $\Psi=\Psi_s$, $H=K=r^2$ $(q^2=0)$, $F(0)\neq 0$.} In this case the rotating solution is ring-singularity free provided
\begin{equation}\label{15b}
F(0)=1,\;F_{,r}(0)=0\quad (F\equiv G).
\end{equation}
This conclusion is easily achieved on Taylor expanding $N$ and $D$ around the point $p_0=(y=0,r=0)$ [Case (2) provides an instance of such expansions].

But under conditions~\eqref{15b} $\lim_{(y,r)\to p_0}R$, which remains finite (compare with~\cite{conf}), does not exist. On the paths $\mathcal{C}_1$ and $\mathcal{C}_2$ through $p_0$ in the $yr$ plane ($y$ axis is horizontal) defined by: $\mathcal{C}_1$: $r=h(y)$ and $h(0)=0$ [where $h_{,y}(0)$ is assumed finite] and $\mathcal{C}_2$: $y=g(r)$ and $g(0)=0$ [$g_{,r}(0)$ is assumed finite] the limits read, respectively
\begin{equation}\label{15c}
\lim_{y\to 0}R=\frac{6h_{,y}(0)^2F_{,rr}(0)}{a^2+h_{,y}(0)^2},\;\lim_{r\to 0}R=\frac{6F_{,rr}(0)}{1+a^2g_{,r}(0)^2}
\end{equation}
which depend on the derivative of $h$ or $g$ and thus do not exist. For instance, the limit on a curve reaching $p_0$ horizontally [$h_{,y}(0)=0$] is zero and that on a curve reaching $p_0$ vertically\footnote{For this type of curves $h_{,y}(0)=\infty$, but we can still use~\eqref{15c} provided we divide numerator and denominator in its r.h.s by $h_{,y}(0)^2$. In general, for any curve reaching $p_0$ no matter how, we have $h_{,y}(0)g_{,r}(0)=1$.} [$g_{,r}(0)=0$] is $6F_{,rr}(0)$.

Notice that the conditions~\eqref{15b} are met by all regular static \bh constructed so far~\cite{regular} and that, not only $F_{,rr}(0)$ is finite, but all derivatives of $F$ are so at $r=0$. Application of this case to regular static \bh allows one to generate all their normal regular rotating counterparts.

The components of the SET [Eqs~\eqref{18a} to~\eqref{18c} or Eqs.~(13) to (14) of~\cite{conf}] remain finite too but do not exist in the limit $(y,r)\to p_0$ where, for instance on the path $\mathcal{C}_1$, we obtain
\begin{align}
&\ep = -\frac{3h_{,y}(0)^4F_{,rr}(0)}{2[a^2+h_{,y}(0)^2]^2},\;p_r=-\ep \nn\\
&p_{\ta}=p_{\phi}=\frac{3h_{,y}(0)^2[2a^2+h_{,y}(0)^2]F_{,rr}(0)}{2[a^2+h_{,y}(0)^2]^2}.
\end{align}

\paragraph*{\textbf{Case (2):} $\Psi=\Psi_g$ ($\Psi\neq\Psi_s$), $H=r^2$, $K(r_0)=0$, $F(r_0)=0$.} In this case $r_{\text{th}}=r_0$. The rotating solution has a ring singularity at the throat. The evaluation of~\eqref{13} and of its derivative with respect to $r$ on the ring $K(r_0)=0$ and $y=0$ leads to conclude that\footnote{We assume that $\Psi_g$ remains finite on the ring, since otherwise $g_{\ta\ta}=-\Psi_g$ would diverge there.}
\begin{equation}\label{15r}
K_{,r}(r_0)=K_{,rr}(r_0)=0
\end{equation}
[with these values and $K(r_0)=0$, \eqref{13r} is satisfied] and results in $\mathcal{R}_{,r}(r_0)=\mathcal{R}_{,rr}(r_0)=0$ where $\mathcal{R}\equiv \sqrt{F/G}$. So, along the path $\mathcal{C}_3$: $r=h(y)+r_0$ and $h(0)=0$ [where $h_{,y}(0)$ is assumed finite] we obtain the Taylor expansions
\begin{equation}\label{te}
N=20a^4r_0^4y^2+\mathcal{O}(y)^3,\,D=2a^4r_0^4\Psi_g(r_0,0)y^4+\mathcal{O}(y)^5.
\end{equation}

For the type of static wormholes discussed by Morris and Thorne~\cite{W1} and Visser~\cite{W2}, where $F=1-b(r)/r$ in \Sd coordinates, the above conditions are limiting cases of the flare-out condition on the shape function $b$ at the throat $r_0$. Since $G$ is never zero for a wormhole (absence of event horizon~\cite{W2}), $F(r_0)=0$ and $\mathcal{R}_{,r}(r_0)=\mathcal{R}_{,rr}(r_0)=0$ conversely imply, besides $b(r_0)=r_0$, $b_{,r}(r_0)=1$ and $b_{,rr}(r_0)=0$. By Eq.~(11.17) of~\cite{W2}, $b_{,r}(r_0)=1$ is a limit value, and Eq.~(11.13) of~\cite{W2} yields, under the same condition, $b_{,rr}(r_0)<0$ (as is clear from Fig.~11.2 of~\cite{W2}), so that $b_{,rr}(r_0)=0$ could be taken as a limit value too.

With that said, the rotating counterparts of Morris and Thorne wormholes that are written in \Sd coordinates ($F=1-b(r)/r$ and $H=r^2$)
\begin{description}
  \item[(1)] are not ring-singularity free if they are limiting cases ($b_{,r}(r_0)=1$) since in this case $R=N/D$ diverges by~\eqref{te} as $1/y^2$ on the ring $K(r_0)=0$ and $y=0$,
  \item[(2)] are not interpreted as fluids in rotational motion about the $z$ axis, with $T^{\mu\nu}$ given by~\eqref{17}, if they are not limiting cases ($b_{,r}(r_0)<1$) since in this case the constraint $b_{,r}(r_0)<1$ would violate~\eqref{15r}.
\end{description}

\paragraph*{\textbf{Case (3):} $\Psi=\Psi_g$, $H=r^2(l)$.} Here $l$ denotes the proper radial distance that is used as the new radial coordinate and $r$ becomes a function $r(l)$~\cite{W2,W2b}. In this case $F(l) \equiv 1$ and $r_{\text{th}}=r_0=\min\{r(l)\}$. Without loss of generality, we choose $l$ such that $r_{\text{th}}=r_0=r(0)$. Using $l$ as the new radial coordinate we can generate the imperfect fluid rotating counterparts of Morris-Thorne type wormholes~\cite{W1}. These are going to be conformal rotating wormholes if they are massive or normal ones if they are massless. Since $K(l)=r^2(l)/\sqrt{G(l)}\neq 0$ ($G$ is finite for a wormhole), the rotating solution has no ring singularity arising from $\ro^2$ (one can always avoid ring singularities arising from $\Psi_g$ by suitably choosing the latter).

In the following we specialize to the case $\Psi=\Psi_s$ [Eq.~\eqref{14} with $l$ being the new radial coordinate]: $\Psi_s=l^2+q^2+a^2y^2$ and $K=l^2+q^2$. From the definition of $K$ [Eq.~\eqref{K}], we obtain $\sqrt{G(l)}= r(l)^2/(l^2+q^2)$, which satisfies the requirement (11.3) of~\cite{W2}: $\lim_{l\to\pm\infty}G(l)=\text{finite }(=1)$, provided
\begin{equation*}
    \lim_{l\to\pm\infty}r(l)/|l|=1.
\end{equation*}
This latter requirement is necessary in order to have an asymptotically flat spatial static geometry~\cite[Eq.~(11.2)]{W2}.

We have thus determined the general form of the static metric ($F,G,H$)$=$($1,r(l)^4/(l^2+q^2)^2,r(l)^2$) yielding an imperfect fluid rotating wormhole, the metric of which reads [Eqs.~\eqref{10a}, \eqref{10b}]
\begin{align}
&\dd s^2 =\Big(1-\frac{2f}{\ro^2}\Big)\dd t^2-\frac{\ro^2}{\De}\,\dd l^2\nn\\
\label{10aa}&+\frac{4af \sin ^2\theta}{\ro^2}\,\dd t\dd \phi-\ro^2\dd \ta^2-\frac{\Si\sin ^2\theta}{\ro^2}\,\dd \phi^2
\end{align}
\begin{align}
\label{10bb}&\dd s^2 =\frac{\De}{\ro^2}(\dd t-a\sin^2\ta\dd \phi)^2-\frac{\ro^2}{\De}\,\dd l^2-\ro^2\dd \ta^2\nn\\
&-\frac{\sin^2\ta}{\ro^2}\,[a\dd t-(K+a^2)\dd \phi]^2
\end{align}
with $\Psi_s=\ro^2=l^2+q^2+a^2\cos^2\ta$, $2f(l)=l^2+q^2-r(l)^2$, $\De(l)= r(l)^2+a^2$ and $\Si = (l^2+q^2+a^2)^2-a^2\De\sin^2\ta$.

The mass of the wormhole is determined by the requirement
\begin{equation}\label{mass1}
    \Big(\frac{\dd r}{\dd l}\Big)^2\simeq 1-\frac{2m}{r}\;\text{ as }\;r\to \infty
\end{equation}
which results in~\cite{W2b}
\begin{equation}\label{mass2}
    r\simeq |l|-m\ln(|l|/r_0)\;\text{ as }\;|l|\to \infty
\end{equation}
on both sheets of the wormhole. Using this in the $tt$-component of the static and rotating metrics, we arrive at, respectively
\begin{align}
\label{tt1}&G\simeq 1-4m\frac{\ln(r/r_0)}{r}\;\text{ as }\;r\to \infty \\
\label{tt2}&1-\frac{2f}{\ro^2}\simeq 1-2m\frac{\ln(r/r_0)}{r}\;\text{ as }\;r\to \infty .
\end{align}
Thus, time runs at the same rate on both sheets of the (static or rotating) wormhole but, to the order $\ln(r/r_0)/r$, it runs at lower rate for the rotating wormhole than for the static one.

Notice that, since in Eq.~\eqref{10aa} the asymptotic expansion of $\De/\ro^2\simeq 1-2m\ln(r/r_0)/r$ does not include a term proportional to $1/r$, the (asymptotic) mass of the rotating wormhole is that of the static one.

Solutions with $r^2=l^2+p^2$ ($p^2>0$) are massless ($m=0$). In this case $2f=q^2-p^2=\text{const.}$ Without loss of generality, we assume $q^2\geq p^2$. The angular velocity $\Om$ of the rotating wormhole~\eqref{10aa} is defined by $g_{\ta\phi}=\Om g_{\ta\ta}\sin^2\ta$ leading to $\Om(r,\ta) =2af/\ro^4$: This is the angular velocity, attributable to dragging effects, of freely falling particles initially at rest at spatial infinity as they reach the point ($r,\ta$). Thus, the massless rotating wormholes~\eqref{10aa} have no dragging effects if $q^2= p^2$. This latter case will be treated in more details in Sect.~\ref{sec3}.

\paragraph*{\textbf{Case (4):} $\Psi$ any solution to~\eqref{13}, \eqref{13r}, $F>0$, $H>0$ for all $r$.} In this case $r$ is not the proper distance (the case where the proper distance is the radial variable is treated in Case (3), so we won't consider it here).

Asymptotic flatness requires $\lim_{r\to\infty}H/r=1$. This case includes Bronnikov-Ellis static wormholes~\cite{BE1,BE2} [$G=F$, $H=(r^2+q^2)/F$, $q^2\neq 0$] as well as some regular \bh among which we find loop \BH~\cite{loop}.  The rotating solution is a regular wormhole or \abh provided $\Psi$ is suitably chosen.

We provide an example from loop \bh (Bronnikov-Ellis static wormholes are treated in more details in Sect.~\ref{sec3}). Consider the metric (2) of~\cite{loop}:
\begin{align}
&F=\frac{r^4(r-r_+)(r-r_-)}{(r+r_*)^2(r^4+a_0^2)},\; \frac{F}{G}=\Big(\frac{r}{r+r_*}\Big)^4\nn\\
\label{loop1}&H=r^2+\frac{a_0^2}{r^2}
\end{align}
where $H_{\text{min}}=2a_0$. Here ($r_-,r_+$) are the two horizons and $a_0$ and $r_*\equiv \sqrt{r_+r_-}$ are constants. From the definition of $K$ we obtain
\begin{equation}\label{loop2}
K(r)=(r^4+a_0^2)/(r+r_*)^2\neq 0.
\end{equation}
Hence, the imperfect fluid rotating loop \abh has no ring singularity.

The rotating loop \abh is given by Eqs.~\eqref{10a}, \eqref{10b}. In this case $\Psi=\Psi_s$ is not a possible solution for $K\neq r^2+q^2$. It is generally possible to perform a coordinate transformation $r\to R(r)$ by which $K$ transforms as: $K\to K=R^2+q^2$ (see~\cite{conf} for an example). If this is the case, $\Psi=\Psi_s=R^2+q^2+a^2y^2$ can be used as a solution for all $R$. We may investigate such a possibility in a subsequent work. In this work, rather, we restrict ourselves to the spatial asymptotic region ($r\to\infty$) and discuss some physical properties of the rotating loop \aBH.

Similarly, $\Psi=\Psi_n$ [Eq.~\eqref{15}] is not a possible solution too for Eq.~\eqref{A8} is not satisfied. It might be possible too that by a coordinate transformation $r\to R(r)$ a solution of the form~\eqref{15} becomes possible.

With that said, the rotating loop \abh is then a conformal fluid. It is possible to investigate most physical properties of these rotating solutions, without fixing $\Psi_g$ [Eq.~\eqref{15rr}], from the properties of the metric inside the square brackets in~\eqref{10a}. We restrict ourselves to the spatial asymptotic region. As $r\to\infty$, $K\to (r-r_*)^2+2r_*^2$. In terms of the new radial coordinate $R=r-r_*$ and $q^2=2r_*^2$, $\Psi_s\simeq R^2+q^2+a^2y^2$ is an asymptotic solution. This is not enough to assert that the conformal rotating fluid behaves asymptotically as a normal one since the inequality $G\neq F$ holds even asymptotically [$G-F =2\sqrt{2}q/R+O(1/R^2)$]: It behaves that way only approximately since $q=\sqrt{2}r_*$ and $r_-$ are close to zero~\cite{loop}, so for very large distances from the source we assume $G\simeq F$.

Asymptotically, the factor $\Psi_s/\ro^2$ in~\eqref{10a} is 1 and its series expansion has no term proportional to $1/R$, so we will drop it. The rotating loop \abh behaves asymptotically as:
\begin{align}
&\dd s^2 \simeq \Big(1-\frac{2m}{R}\Big)\dd t^2-\frac{1}{1-\frac{2m}{R}}\,\dd R^2\nn\\
\label{loop3}&+\frac{4ma \sin ^2\theta}{R}\,\dd t\dd \phi-R^2\dd \ta^2-R^2\sin ^2\theta\,\dd \phi^2
\end{align}
where we have used the definitions of $\ro^2$, $2f(r)$, $\De$ and $\Si$ given in the sentence preceding~\eqref{10a} along with~\eqref{loop1} and~\eqref{loop2}. Here $ma$ is the angular momentum and $m=(r_++r_-)/2$ is the mass of the rotating loop \aBH, which is slightly lower than that of the static loop one given by $m_{\text{stat}}=m+q/\sqrt{2}$ and slightly larger than that of the Kerr solution $m_{\text{Kerr}}=r_+/2<m$.

\section{Rotating imperfect fluid wormholes \label{sec3}}

In the following we assume that the static solution~\eqref{1} is a wormhole solution. We keep on doing general treatments and we won't fix the form of any metric component of~\eqref{1}, nor shall we fix the function $\Psi$ in~\eqref{10}, until we consider specific applications.

We consider a static wormhole of the Bronnikov-Ellis type with $G(r)=F(r)$ and $H=(r^2+q^2)/F$ where we take $q^2>0$
\begin{equation}\label{s1}
\dd s_{\text{stat}}^2 = F\dd t^2 - \frac{\dd r^2}{F} - \frac{(r^2+q^2)}{F}\,(\dd \ta^2+\sin^2\ta\dd \varphi^2),
\end{equation}
where, in this case, $K=H$ and $\ro^2=H+a^2y^2$. The radius of the static throat $r_0$ is the minimum value of $\sqrt{H}$ which occurs at $r_{\text{th}}$: $r_0=\sqrt{2r_{\text{th}}/F_{,r}(r_{\text{th}})}$ if $F\neq 1$ or $r_0=|q|$ if $F=1$.

The angular velocity $\Om$ of the rotating wormhole~\eqref{10} is defined by $g_{\ta\phi}=\Om g_{\ta\ta}\sin^2\ta$ leading to $\Om(r,\ta) =2af/\ro^4$: This is the angular velocity, attributable to dragging effects, of freely falling particles initially at rest at spatial infinity as they reach the point ($r,\ta$). Assuming asymptotic flatness of the static wormhole: $F=1-2mr^{-1}+\mathcal{O}(r^{-2})$, then $\Om\to 2Jr^{-3}$ as $r\to\infty$ where $J=ma$ is the angular momentum of the rotating wormhole and $m$ is the mass of the static one. The angular velocity of the particles of the rotating exotic fluid~\cite{Magli,conf} as they pass by the point ($r,\ta$) is $a/(H+a^2)$, which is given by~\eqref{16} where $e^{\mu}_t$ is the 4-velocity vector of the fluid. $\Om$ is different from $a/(H+a^2)$, this is because the fluid particles do not follow geodesic motion~\cite{conf}. Similarly to rotating \BH, we can define the angular velocity of the throat by $\Om_0\equiv \Om(r_{\text{th}},\pi/2)=a[1-F(r_{\text{th}})]/r_0^2$ and its linear velocity by $\Om_0 r_0$.

The rotating massless wormhole, where $F=1$, $m=0$, $f\equiv 0$, $\Om\equiv 0$, has thus no dragging effects: Its particles rotate with the angular velocity $a/(r^2+q^2+a^2)$ but the freely falling particles do not acquire any angular velocity.

Now, we want to evaluate the effects of rotation on the mass and conditions of traversability. It is obvious from~\eqref{10a} that if $G=F$, $F\to 1-2mr^{-1}$ and $H\to r^2$ as $r\to\infty$ with $\Psi\to H$ as $a\to 0$ (being normal), then $g_{tt}\to 1-2mr^{-1}$ as $r\to\infty$. Thus, rotation has no effect on the mass of the rotating wormhole. An early work on slowly rotating wormholes concluded that the mass of the rotating wormhole increases with rotation~\cite{sec}. The discrepancy resides in our choice of the source term $T^{\mu\nu}$ being that of a fluid having only a rotational motion about a fixed axis (here $Oz$ with $G_{r\ta}\equiv 0$) while for the source term of~\cite{sec,ks}, where $G_{r\ta}\neq 0$, Eq.~\eqref{17} no longer holds. Moreover, in~\cite{sec,ks} the extra condition $T_{\phi}{}^t=0$ was used. Had we imposed the same condition we would have obtained, using~\eqref{10}, \eqref{16} and~\eqref{17}, $T_{\phi}{}^t=-a\sin^2\ta (H+a^2)(\ep +p_{\phi})/\ro^2=0$ leading to $p_{\phi}=-\ep$ so that our fluid is no longer totally imperfect. More on conditions to have fluid solutions are found in~\cite{W3,W4,W4b}.

If the static wormhole is traversable, then this property is generally not altered by rotation but changes to the specifications of the conditions of traversability necessarily occur due to dragging effects. We won't elaborate any more on this point.

Since they are based solely on the general form of the rotating metric~\eqref{10a} ($\Psi$ not fixed), all the above conclusions made in this section do not depend on the way one interprets the source term $T^{\mu\nu}$. In the following, we focus on two different interpretations and restrict ourselves to the massless case $m=0$ taking $\Psi=\Psi_s=\ro^2$ ($q^2> 0$) since it is the unique solution in this case (see appendix).

\subsection{Rotating imperfect fluid wormhole without electromagnetic field \label{sec3a}}

If $m=0$ then $F=1$. Here we assume that the source term $T^{\mu\nu}$ constitutes an imperfect exotic fluid given by~\eqref{17} to~\eqref{18c} [since $\Psi=\Psi_s=\ro^2$ and $G=F$, it would be better to use Eqs.~(13) and (14) of~\cite{conf}]. We find
\begin{multline}\label{d1}
T^{\mu\nu}=-\frac{q^2}{\ro^4}\big[1+\frac{2a^2\sin^2\ta}{\ro^2}\big] e^{\mu}_te^{\nu}_t-\frac{q^2}{\ro^4}e^{\mu}_re^{\nu}_r\\
+\frac{q^2}{\ro^4}e^{\mu}_{\ta}e^{\nu}_{\ta}
+\frac{q^2}{\ro^4}\big[1+\frac{2a^2\sin^2\ta}{\ro^2}\big]e^{\mu}_{\phi}e^{\nu}_{\phi}
\end{multline}
where in this case $\ro^2=r^2+q^2+a^2\cos^2\ta$. The basis $(e_t,\,e_r,\,e_{\ta},\,e_{\phi})$ and the rotating metric are given by~\eqref{16} and~\eqref{10a}, respectively, with $K=H=r^2+q^2$, $f=0$ and $\De =r^2+q^2+a^2$:
\begin{equation}\label{d2}
\dd s^2 =\dd t^2-\frac{\ro^2}{\De}\dd r^2-\ro^2\dd \ta^2-\De\sin ^2\theta\dd \phi^2.
\end{equation}

We proceed now to compare the exotic matter content of the rotating imperfect exotic fluid wormhole $|\ep|$ with that of the static one $|\ep_{\text{st}}|$. The static wormhole counterpart of~\eqref{d2}, the metric of which is obtained from~\eqref{d2} setting $a=0$ or from~\eqref{s1} setting $F=1$, is a perfect fluid with a negative density and isotropic pressure, its SET is given by
\begin{equation}\label{s3}
    T_{\text{st}}^{\mu\nu}=\frac{q^2}{(r^2+q^2)^2}{\rm diag}(-1,-1,1,1).
\end{equation}
From~\eqref{d1} and~\eqref{s3} we have respectively
\begin{equation*}
 |\ep|=\frac{q^2(H+2a^2-a^2y^2)}{(H+a^2y^2)^3},\;|\ep_{\text{st}}|=\frac{q^2}{H^2}.
\end{equation*}
It is obvious that, for fixed ($r,q,a$), $|\ep|$ decreases with increasing $y^2$. Moreover, $|\ep|(y^2=1)=q^2/(H+a^2)^2<|\ep_{\text{st}}|$. This implies the existence of a minimum value $y_{\text{min}}^2$ beyond which $|\ep|<|\ep_{\text{st}}|$.

The minimum value $y_{\text{min}}^2$ is a function of $a^2$, solution to $a^4y_{\text{min}}^6+3Ha^2y_{\text{min}}^4+4H^2y_{\text{min}}^2-2H^2=0$. Without solving the latter equation, it is easy to see that in the limit $a^2\to \infty$, we have $y_{\text{min}}^2\to 0$ and that by differentiation ($r$ and $q$ are held constant) we have $\dd y_{\text{min}}/\dd a<0$. This shows that the exotic matter required to hold the rotating imperfect exotic fluid wormhole is less than that of its static counterpart and becomes much smaller with rotation.

\paragraph*{\textbf{Stability issues.}} Axial perturbations of static wormholes with the above structure of $T_{\text{st}}^{\mu\nu}$ [Eq.~\eqref{s3}], without electromagnetic field, were included in the investigation carried on in~\cite{st0}. Schr\"{o}dinger-like Eqs.~(32) and~(33) of~\cite{st0}, where $H_2(r)$ is a radial gravitational perturbation, $V_{\text{eff}}(r)$ is the effective Schr\"{o}dinger potential, and $\om$ is the frequency of oscillations coming from the factor $\e^{\ii \om t}$ used to proceed to the separation of the time variable, apply to our static wormhole. In the case of the perfect fluid static wormhole, $V_{\text{eff}}(r)$ reads
\begin{equation}\label{s2}
\hspace{-1mm}V_{\text{eff}}=\frac{[(\ell+2)(\ell-1)+2]r^2+[(\ell+2)(\ell-1)-1]q^2}{(r^2+q^2)^2}
\end{equation}
where $\ell$ is the multi-pole order. In his book on the mathematical theory of \BH, Chandrasekhar has ignored the case $\ell=1$ when dealing with both axial and polar perturbations of the \Sd \aBH~\cite[chap. 4, \S 24]{mtbh}, thus considering the quadruple excitation ($\ell=2$) as the leading dynamical gravitational order.

Now, it is straightforward to check that the expression~\eqref{s2} of $V_{\text{eff}}(r)$ is positive definite for all $\ell\geq 2$, which is a sufficient condition for the existence of asymptotically well-behaved oscillating solutions, that is, solutions with positive squared frequencies $\om^2>0$. We thus conclude to the existence of stable modes of axial perturbations of the perfect fluid static wormhole with the above structure of $T_{\text{st}}^{\mu\nu}$. We also conclude to the stability against all relevant dynamical axial perturbations ($\ell\geq 2$).

Concerning the stability of the imperfect fluid rotating massless wormhole, without electromagnetic field, against small perturbations, we extend the above-made conclusion and conjecture that the rotating counterpart wormhole [where $T^{\mu\nu}$ is given by~\eqref{d1}] of the static background one [where $T_{\text{st}}^{\mu\nu}$ is given by~\eqref{s3}] is stable against linear axial perturbations. This statement is at least true for small values of the rotation parameter $a$.

\subsection{Rotating imperfect fluid wormhole with electromagnetic field \label{sec3b}}

Very recently, Bronnikov et al.~\cite{ex}, and references therein, reinterpreted the source term $T_{\text{st}}^{\mu\nu}$ of a massless Ellis static wormhole as being due to two contributions $T_{\text{st}}^{\mu\nu}=T_{em-\text{st}}^{\mu\nu}+T_{d-\text{st}}^{\mu\nu}$ where $T_{em-\text{st}}^{\mu\nu}$ is attributable to a source-free radial electric or magnetic field and $T_{d-\text{st}}^{\mu\nu}$ is that of a perfect fluid (pressureless dust) with negative density
\begin{align}
\label{d3}&T_{em-\text{st}}^{\mu\nu}=\frac{q^2}{(r^2+q^2)^2}\,{\rm diag}(1,-1,1,1)\\
\label{d4}&T_{d-\text{st}}^{\mu\nu}=-\frac{2q^2}{(r^2+q^2)^2}u^{\mu}u^{\nu},\, [u^{\mu}=(1,0,0,0,)].
\end{align}
satisfying~\eqref{s3} = \eqref{d3} + \eqref{d4}.

When the wormhole rotates none of the above two components remains diagonal; because of the motion, besides the basis~\eqref{16} which rotates with the fluid, the SET of the windy dust acquires a $\phi\phi$-component due to the pressure in the $e^{\mu}_{\phi}$ direction, so that it no longer represents a perfect fluid. The total $T^{\mu\nu}$ which now splits as $T^{\mu\nu}=T_{em}^{\mu\nu}+T_{d}^{\mu\nu}$ is still given by~\eqref{d1} with
\begin{align}
\label{d5}&T_{em}^{\mu\nu}=\frac{q^2}{\ro^4}[e^{\mu}_te^{\nu}_t-e^{\mu}_re^{\nu}_r
+e^{\mu}_{\ta}e^{\nu}_{\ta}+e^{\mu}_{\phi}e^{\nu}_{\phi}] \\
\label{d6}&T_{d}^{\mu\nu}=\frac{2q^2}{\ro^6}
[-\De \,e^{\mu}_te^{\nu}_t+a^2\sin^2\ta \,e^{\mu}_{\phi}e^{\nu}_{\phi}]
\end{align}
($\De =r^2+q^2+a^2$) which reduce to~\eqref{d3} and~\eqref{d4} if rotation is suppressed. The metric is still given by~\eqref{d2}.

The exotic matter required to hold this rotating wormhole, with electromagnetic filed, is less than that of its static counterpart. From~\eqref{d6} and~\eqref{d4} we have $|\ep_d|=2q^2 (H+a^2)/(H+a^2y^2)^3$ is smaller than $|\ep_{d-\text{st}}|=2q^2/H^2$ if $y^2>y_{\text{min}}^2\equiv [(H^3+a^2H^2)^{1/3}-H]/a^2$, where $y_{\text{min}}^2<1/3$ and $y_{\text{min}}^2\to 0$ as $a^2\to\infty$, and becomes much smaller with rotation.

\paragraph*{\textbf{Stability issues.}} As is well known, the stability analysis depends on the matter components making up the SET. The stability analysis of the metric~\eqref{s1}, with the SET split as a sum of a source-free radial electric or magnetic field $T_{em-\text{st}}^{\mu\nu}$ and a perfect fluid (pressureless dust) with negative density $T_{d-\text{st}}^{\mu\nu}$, has been investigated in a couple of papers~\cite{st1,st2,st3} and recently in~\cite{ex}. The analysis made in~\cite{ex} completes and generalizes that of~\cite{st2}.

It was shown that if $T_{\text{st}}^{\mu\nu}=T_{em-\text{st}}^{\mu\nu}+T_{d-\text{st}}^{\mu\nu}$, then the model admits stable as well as unstable modes depending on how the background static wormhole is perturbed. Moreover, within the polar mode of perturbation, while the analysis made in~\cite{ex} has completed that of~\cite{st2}, however it concerned only with the case where the equation of state obeys some power-law formula ensuring positiveness of the potential function in the master equation governing the dynamics of the perturbations. No physical argument was given as to why such choice of the equation of state. The question of stability remains thus open to other choices of the equation of state and to cases where the positiveness of the potential is not ensured.

In such a situation one should conclude to the instability of the model~\cite{st-hete} since if the background static wormhole is ``abandoned" to itself, one a priori does not know in which direction would evolve the initial perturbations as there is no control parameter on which one acts to drive the evolution.

Concerning axial perturbations, the situation is quit different, in that, no special choice of whatever perturbation function was made, and thus the conclusion to the stability against linear axial perturbation is general~\cite{ex}.

Concerning the stability of the imperfect fluid rotating massless wormhole, with electromagnetic field, against small perturbations, we may extend the conclusions made in~\cite{ex} and conjecture that the rotating counterpart wormhole (where $T^{\mu\nu}=T_{em}^{\mu\nu}+T_{d}^{\mu\nu}$ still holds) of the static background one (with $T_{\text{st}}^{\mu\nu}=T_{em-\text{st}}^{\mu\nu}+T_{d-\text{st}}^{\mu\nu}$) is stable against linear axial perturbations. This statement is at least true for small values of the rotation parameter $a$. This statement does not exclude the existence of unstable modes due to different ways of perturbations, as is the case with the static background wormhole.

\section{Conclusion \label{sec4}}

We have derived a shortcut formula for generating rotating metrics. The metric formula appears to be very useful in that the rotating solution acquires the properties of a fluid in rotational motion about a fixed axis if the rotating-metric component $g_{\ta\ta}=-\Psi$ obeys two given differential equations one of which is nonlinear.

Moreover, given a static metric one may derive different rotating solutions depending on the form of the function $\Psi$. Conversely, given two equivalent (related by a coordinate transformation) static metrics, the shortcut metric formula does indeed generate two imperfect, however, non-equivalent rotating fluid solutions using the same $\Psi$~\cite{conf}. As a consequence of that, the generated rotating solution from a Morris-Thorne type static wormhole in \Sd coordinates is not always a regular solution or a fluid one. This last property has the advantage that by a coordinate transformation on the radial coordinate one can modify the forms of $F$ and $H$ to get the desired rotating metric (see~\cite{conf} for further illustrated examples).

We have shown that regular static \bh with $g_{tt}g_{rr}= 1$ ($g_{\ta\ta}=-(r^2+q^2)$, $q^2\geq 0$) have their rotating counterparts regular too as they are the rotating counterparts of Morris and Thorne wormholes in non-\Sd coordinates where the radial coordinate is the spatial proper distance. We have also concluded if Morris and Thorne static wormholes are written in \Sd coordinates then their rotating counterparts are neither regular solutions nor fluids obeying the constraints $G_{\mu\nu}=T_{\mu\nu}$ where $T_{\mu\nu}$ is an imperfect fluid given by~\eqref{17}.

If the exotic matter sustaining the throat is modeled by a fluid, in our case a \textit{totally imperfect} one, then the rotation has no effect on the mass of the wormhole nor does it affect much the conditions of traversability provided the dragging effects do not accelerate freely falling objects beyond Earth's gravity acceleration. The energy of rotation of the wormhole is communicated to the fluid particles, which each rotates with an angular velocity of $a/(K+a^2)$, keeping the mass of the wormhole invariant.

We have briefly discussed an imperfect fluid rotating loop \abh and shown how its mass tinily exceeds that of a Kerr solution with the same event horizon $r_+$.

We have derived the rotating counterpart of the stable exotic dust Ellis wormhole emerged in a source-free radial electric or magnetic field. In all cases the rotating massless wormhole has not dragging effects. Stabilities issues were also discussed, generalizing the results made in~\cite{ex} we have concluded to the stability against small axial perturbations.

Other suggested metrics~\cite{teo} for generating rotating wormholes, used also in~\cite{W4,W5}, failed to generate fluid wormholes~\cite{W3}. Such metrics, where $g_{\phi\phi}(r,\ta)/g_{\ta\ta}(r,\ta)\equiv \sin^2\ta$, cannot be brought to the form~\eqref{10b}. It has been shown that the source term for such generated rotating wormholes, found in~\cite{teo}, is not that of a fluid~\cite{W3,W4}. However, the elements of the proof given in~\cite{W3} rely on the assumption that the fluid undergoes only a rotational motion about a fixed axis. So, it might still be possible to attach a fluid interpretation to the general metric generating rotating wormholes~\cite{teo} (but not to the specific example of Teo wormhole~\cite{teo} as it violates the condition $G_{r\ta}=0$) if (1) one considers, besides the rotational motion, a radial motion too, and (2) one imposes the condition $G_{r\ta}=0$ which constraints the components of Teo general rotating metric.

In subsequent works we will extend the analysis to include other static wormhole solutions~\cite{an1}, among which we have wormholes in Wyman's solution~\cite{an2} and wormholes in Ho\v{r}ava theory~\cite{an3}, and we will generate their imperfect fluid rotating counterparts.

\section*{\large Appendix: Proofs of uniqueness}
\renewcommand{\theequation}{A.\arabic{equation}}
\setcounter{equation}{0}

\paragraph*{\textbf{Step 1.}} We intend to show that if $K=H=r^2+q^2$ (in this case $G=F$) and $q^2\neq 0$, then the unique solution to the system~\eqref{13} and~\eqref{13r} is $\Psi=\text{constant }(r^2+q^2+a^2y^2)$. This will prove the uniqueness of the rotating solutions generated in Sects.~\ref{sec3a} and~\ref{sec3b}. It is more convenient to use the general form~\eqref{15}: $\Psi=H\exp{[a^2\psi(r,y^2,a^2)]}$ without assuming that $\lim_{a\to 0}\Psi=H$. Transforming to the coordinates $r\to r$, $y^2\to x=r^2y^2$, by which the derivatives transform as $\Psi_{,r}\to \Psi_{,r}+2(x/r)\Psi_{,x}$ and $\Psi_{,y^2}\to r^2\Psi_{,x}$ (same transformations for the derivatives of $\psi$), Eq.~\eqref{13r} becomes
\begin{equation}\label{A1}
\psi_{,r}=-\frac{2(2r^2+q^2)x}{r(r^2+q^2)(r^4+q^2r^2+a^2x)}
\end{equation}
yielding the solution
\begin{align}
\label{A2}&a^2\psi=\ln \Big[\frac{r^2+q^2+a^2 y^2}{r^2+q^2}\Big]+a^2g(x) \\
\label{A3}&\Psi=(r^2+q^2+a^2 y^2)f(x)
\end{align}
where $f(x)=\exp{[a^2g(x)]}$ are any functions of $x$. Now, injecting~\eqref{A3} into~\eqref{13} we reduce it to
\begin{equation}\label{A4}
x(\ro^2 +q^2)(3f_{,x}^{\ \; 2}-2ff_{,xx})-(\ro^2 +2q^2)ff_{,x}=0
\end{equation}
where in this case $\ro^2=r^2+a^2 y^2$. Differentiating~\eqref{A4} two times with respect to $r$ we obtain
\begin{equation}\label{A5}
    x(3f_{,x}^{\ \; 2}-2ff_{,xx})-ff_{,x}=0,
\end{equation}
which we insert back in~\eqref{A4} to eliminate $f_{,xx}$, the remaining equation reads
\begin{equation}\label{A6}
    q^2ff_{,x}=0
\end{equation}
resulting in $f=\text{constant}$ if $q^2\neq 0$. If $q^2=0$, Eq.~\eqref{A4} is consistent with~\eqref{A5} leading to, besides the trivial solution $f=\text{constant}$, $f(x)=c_1/(\sqrt{x}+c_2)^2$, where $c_1$, $c_2$ are constants and
\begin{equation}\label{A7}
    \Psi=\frac{c_1(r^2+a^2 y^2)}{(r|y|+c_2)^2}.
\end{equation}
Notice that this last solution is not of the form~\eqref{15} since $\lim_{a\to 0}a^2\psi=\ln [c_1/(r|y|+c_2)^2]$ so that it does not have a Taylor series in powers of $a^2$ of the form $\Psi_n=H+\sum_{i=1}a^{2i}X_{2i}(r,y^2)$ where the first (independent) term is $H=r^2$.

\paragraph*{\textbf{Step 2.}} Now, we intend to prove that if $G=F$ and if $\Psi$ has a Taylor series in powers of $a^2$, then $\Psi_s$ is the only solution of form~\eqref{15}. If $G\neq F$, other solutions of form~\eqref{15} are possible. Keeping the two first terms of the series, $\Psi_n=H+a^{2}X_{2}(r,y^2)+\cdots$, Eqs.~\eqref{13}, \eqref{13r} result in three leading equations which we combine to build the following simplified expressions (we do not assume yet $G=F$):
\begin{align}
\label{A8}&K H_{,r} K_{,r}-H K_{,r}^{\ \; 2}+H K (K_{,rr}-2)=0\\
\label{A9}&X_2=\frac{H^2 (8 K-K_{,r}^{\ \; 2}) y^2}{K^2 (8 H-H_{,r} K_{,r})}\\
&K_{,r} (8 K-K_{,r}^{\ \; 2}) K_{,rrr}+K_{,r}^{\ \; 2} (K_{,rr}-2)^2\nn\\
\label{A10}&-4 K K_{,rr} (K_{,rr}+4)+48 K=0.
\end{align}
Eq.~\eqref{A8} provides $H$ in terms of $K$ by integration
\begin{equation}\label{A11}
H=c \exp \Big[\int ^r\frac{K_{,z}^{\ \; 2}-K(z) (K_{,zz}-2)}{K_{,z} K(z)}\,\dd z\Big]
\end{equation}
where $c$ is a constant. If $G=F$, then $K=H$ and Eq.~\eqref{A8} yields $H=r^2+q^2=K$ ($q^2\neq 0$), and by Step 1, $\Psi$ (rather the Taylor series of $\Psi$) reduces to $\Psi_s$. If $G\neq F$, then Eqs.~\eqref{A8}, \eqref{A9} and~\eqref{A10} provide a solution of the form~\eqref{15}.



\begin{thebibliography}{99}

\bb{as1}J.B.~Hartle and D.H.~Sharp, Astrophys. J. \textbf{147}, 317 (1967).

\bb{as2}J.B.~Hartle, Astrophys. J. \textbf{150}, 1005 (1967).

\bibitem{p}A.~Papapetrou, Proc. Roy. Irish Acad. \textbf{52}, {11} (1948).

\bibitem{khat}V.M.~Khatsymovsky, Phys. Lett. B \textbf{429}, {254} (1998).

\bibitem{ks}P.E.~Kashargin and S.V.~Sushkov, Grav. Cosmol. \textbf{14}, {80} (2008), arXiv:0710.5656.

\bibitem{gh}M.~Azreg-A\"{\i}nou, Gen. Relativ. Gravit. \textbf{44}, 2299 (2012), arXiv:1206.1408.

\bibitem{Van} W.~Van Stockum, 1937, Proc. Roy. Soc. Edinb., \textbf{57}, 135 (1937).

\bibitem{Kerr} R.P.~Kerr, Phys. Rev. Lett. \textbf{11}, 237 (1963).

\bibitem{Islam} J.N.~Islam, \textit{Rotating Fields in General Relativity} (Cambridge University Press, Cambridge, 1985).

\bibitem{rot1}A.~Krasinski, J. Math. Phys. \textbf{39}, 2148 (1998), arXiv:gr-qc/9707021.

\bibitem{rot2}B.V.~Ivanov, Class. Quantum Grav. \textbf{19}, 5131 (2002), arXiv:gr-qc/0207013.

\bibitem{Kramer}H.~Stephani, D.~Kramer, M.A.H.~MacCallum, C.~Hoenselaers, and E.~Herlt, \textit{Exact Solutions of Einstein's Field Equations} (Cambridge University Press, Cambridge, 2003).

\bibitem{Herrera}L.~Herrera and J.~Jim\'{e}nez, J. Math. Phys. \textbf{23}, 2339 (1982).

\bibitem{IrinaD}I.~Dymnikova, Gen. Relativ. Gravit. \textbf{24}, 235 (1992).

\bibitem{Magli} A.~Burinskii, E.~Elizalde, S.R.~Hildebrandt, and G.~Magli, Phys. Rev. D \textbf{65}, 064039 (2002), arXiv:gr-qc/0109085.

\bibitem{Via}S.~Viaggiu, Int. J. Mod. Phys. D \textbf{15}, 1441 (2006), arXiv:gr-qc/0603036.

\bibitem{Lemos}J.P.S.~Lemos and V.T.~Zanchin, Phys. Rev. D \textbf{83}, 124005 (2011), arXiv:1104.4790.

\bibitem{Carter}B.~Carter, Commun. Math. Phys. \textbf{10}, 280 (1968).

\bibitem{Ernst}F.J.~Ernst, Phys. Rev. \textbf{167}, 1175 (1968).

\bibitem{Islam1}J.N.~Islam, Proc. R. Soc. Lond. A \textbf{367}, 271 (1979).

\bibitem{Bonnor} W.B.~Bonnor, J. Phys. A: Math. Gen., \textbf{13}, 3465 (1980).

\bibitem{Clement} G.~Cl\'ement, Phys. Rev. D \textbf{57}, 4885 (1998), arXiv:gr-qc/9710109.

\bibitem{plus} E.N.~Glass and J.P.~Krisch, Class. Quantum Grav. \textbf{21}, 5543 (2004), arXiv:gr-qc/0410089.

\bibitem{Gibbons} G.W.~Gibbons, H.~L\"{u}, D.N.~Page and C.N.~Pope, J. Geom. Phys. \textbf{53}, 49 (2005), arXiv:hep-th/0404008.

\bibitem{phantom1}M.~Azreg-A\"{\i}nou, G.~Cl\'ement, J.C.~Fabris, and
M.E.~Rodrigues, Phys. Rev. D \textbf{83} 124001 (2011), arXiv:1102.4093.

\bibitem{Janis1} E.T.~Newman and A.I.~Janis, J. Math. Phys. \textbf{6}, 915 (1965).

\bibitem{the1}M.~Demia\'{n}ski and E.T.~Newman, Bull. Acad. Polon. Sci \textbf{14}, 653 (1966).

\bibitem{the2}M.~Demia\'{n}ski, Phys. Lett. A \textbf{42}, 157 (1972).

\bibitem{Gurses} M.~G\"{u}rses and F.~G\"{u}rsey, J. Math. Phys. \textbf{16}, 2385 (1975).

\bibitem{Janis2} S.P.~Drake and R.~Turolla, Class. Quantum Grav. \textbf{14}, 1883 (1997), arXiv:gr-qc/9703084.

\bibitem{teo}E.~Teo, Phys. Rev. D \textbf{58}, 024014 (1998), arXiv:gr-qc/9803098.

\bibitem{Janis3} S.P.~Drake and P.~Szekeres, Gen. Relativ. Gravit. \textbf{32}, 445 (2000), arXiv:gr-qc/9807001.

\bibitem{Janis4} O.~Brauer, H.A.~Camargo and M.~Socolovsky,
\emph{Newman-Janis Algorithm Revisited}, arXiv:1404.1949.

\bibitem{kr}F.~Caravelli and L.~Modesto, Class. Quantum Grav. \textbf{27}, 245022 (2010), arXiv:1006.0232.

\bibitem{az2}M.~Azreg-A\"{\i}nou, Class. Quantum Grav. \textbf{28}, 148001 (2011), arXiv:1106.0970.

\bibitem{conf} M.~Azreg-A\"{\i}nou, Phys. Lett. B \textbf{730}, 95 (2014), arXiv:1401.0787.

\bibitem{regular}J.M.~Bardeen, in: Proceedings of GR5, Tbilisi, USSR, (1968).\\
E.~Ay\'on--Beato and A.~Garc\'{\i}a, Phys. Lett. B \textbf{464}, 25 (1999), arXiv:hep-th/9911174.\\
A.~Burinskii and S.R.~Hildebrandt, Phys. Rev. D \textbf{65}, 104017 (2002), arXiv:hep-th/0202066.\\
S.A.~Hayward, Phys. Rev. Lett. \textbf{96}, 031103 (2006), gr-qc/0506126.\\
W.~Berej, J.~Matyjasek, D.~Tryniecki, and M.~Woronowicz, Gen. Relativ. Gravit. \textbf{38}, 885 (2006), arXiv:hep-th/0606185.

\bibitem{W1}M.S.~Morris and K.S.~Thorne, Am. J. Phys. \textbf{56}, 395 (1988).

\bibitem{W2}M.~Visser, \textit{Lorentzian wormholes: from Einstein to Hawking} (AIP Press, Cambridge, 1995).

\bibitem{W2b}M.S.~Morris, K.S.~Thorne, and U.~Yurtsever Phys. Rev. Letts \textbf{61}, 1446 (1988).

\bibitem{BE1}K.A.~Bronnikov, Acta Phys. Pol. B \textbf{4}, 251 (1973).

\bibitem{BE2}H.G.~Ellis, J. Math. Phys. \textbf{14}, 104 (1973).

\bibitem{loop}E.~Alesci and L.~Modesto, Gen. Relativ. Gravit. \textbf{46}, 1656 (2014), arXiv:1101.5792.

\bibitem{sec}P.E.~Kashargin and S.V.~Sushkov, Phys. Rev. D \textbf{78}, 064071 (2008), arXiv:0809.1923.

\bibitem{W3}S.E.~Perez Bergliaffa and K.E.~Hibberd, arXiv:gr-qc/0006041.

\bibitem{W4}P.K.F.~Kuhfittig, Phys. Rev. D \textbf{67}, 064015 (2003), arXiv:gr-qc/0401028.

\bibitem{W4b}M.~Ishak and K.~Lake, Phys. Rev. D \textbf{68}, 104031 (2003), arXiv:gr-qc/0304065.

\bibitem{st0}K.A.~Bronnikov, R.A.~Konoplya, A.~Zhidenko, Phys. Rev. D \textbf{86}, 024028 (2012), arXiv:1205.2224.

\bibitem{mtbh}S.~Chandrasekhar, \textit{The Mathematical Theory of Black Holes} ((Clarendon Press, Oxford, 1998).

\bibitem{ex}K.A.~Bronnikov, L.N.~Lipatova, I.D.~Novikov, and A.A.~Shatskiy, Grav. Cosmol. \textbf{19}, 269 (2013), arXiv:1312.6929.

\bibitem{st1}D.I.~Novikov, A.G.~Doroshkevich, I.D.~Novikov, and A.A.~Shatskii, Astron. Rep. \textbf{53}, 1079 (2009).

\bibitem{st2}I.~Novikov, A.~Shatskiy,
JETP \textbf{114} (5), 801-804 (2012), arXiv:1201.4112.

\bibitem{st3}O.~Sarbach, T.~Zannias, Phys. Rev. D \textbf{81}, 047502 (2010), arXiv:1001.1202.

\bibitem{st-hete}M.~Azreg-A\"{\i}nou, Class. Quantum Grav. \textbf{16}, 245 (1999), arXiv:gr-qc/9902005.\\
M.~Azreg-A\"{\i}nou, G.~Cl\'ement, C.P.~Constantinidis, and J.C.~Fabris, Grav. Cosmol. \textbf{6}, 207 (2000), arXiv:gr-qc/9911107.

\bibitem{W5}F.S.N.~Lobo, in: Classical and Quantum Gravity Research, pp. 1-78, Eds. M.N.~Christiansen and T.K.~Rasmussen, (Nova Science Publishers, N.Y. 2008), arXiv:0710.4474.

\bibitem{an1}M.~Hohmann, Phys. Rev. D \textbf{89}, 087503 (2014), arXiv:1312.5290.

\bibitem{an2} J.B.~Formiga and T.S.~Almeida,
\emph{Wormholes in Wyman's solution}, arXiv:1404.0328.

\bibitem{an3} J.~Bellorin, A.~Restuccia and A.~Sotomayor,
\emph{Wormholes and naked singularities in the complete Ho\v{r}ava theory}, arXiv:1404.2884.

\end{thebibliography}
\end{document}